\documentclass[
	paper=a4,
	DIV=27,+-++++
	fontsize=11pt	
	]{scrartcl}

	\KOMAoption{captions}{tableheading, figuresignature, nooneline, bottombeside, outerbeside}
	\addtokomafont{caption}{\small}			
	\addtokomafont{captionlabel}{\bfseries} 
	
		 \addtokomafont{author}{\small}
		 \addtokomafont{date}{\small}
		 \addtokomafont{publishers}{\small}

\usepackage[utf8]{inputenc} 
\usepackage[T1]{fontenc}
\usepackage[english]{babel}
\usepackage{ulem} 
\usepackage[sfdefault, light, condensed]{roboto}	    
\usepackage{amsmath}				
\usepackage{amssymb}				
\usepackage{siunitx}	            
	\DeclareSIUnit{\angstrom}{\textup{\AA}}

\usepackage[version=4]{mhchem}      

\usepackage[onehalfspacing]{setspace} 
\usepackage{graphicx}
	\graphicspath{ {images/} }

\usepackage{authblk}	
\usepackage{textgreek}	
    \usepackage{upgreek}

\usepackage[dvipsnames]{xcolor}

\usepackage{hyperref}

\usepackage[sort&compress,square,numbers,comma]{natbib}			

\title{Reversible Switching of the Environment-Protected Quantum Spin Hall Insulator Bismuthene at the Graphene/SiC Interface}
    \author[1,2]{Niclas Tilgner}
    \author[1,2]{Susanne Wolff}
    \author[3,4]{Serguei Soubatch} 
    \author[5]{Tien-Lin Lee}
    \author[1]{Andres David Pe\~{n}a Unigarro}
    \author[1]{Sibylle Gemming}
    \author[3,4,6]{F. Stefan Tautz}
    \author[3,4,6,$\ddagger$]{Christian Kumpf}
    \author[1,2]{Thomas Seyller}
    \author[1,2,$\dagger$]{Fabian G\"{o}hler}
    \author[1,2,*]{Philip Sch\"{a}dlich}
	
    \affil[1]{Institute of Physics, Chemnitz University of Technology, 09126 Chemnitz, Germany}
    \affil[2]{Center for Materials, Architectures and Integration of Nanomembranes (MAIN), 09126 Chemnitz, Germany}
    \affil[3]{Peter Gr\"{u}nberg Institut (PGI-3), Forschungszentrum J\"{u}lich, 52425 J\"{u}lich, Germany}
    \affil[4]{J\"{u}lich Aachen Research Alliance (JARA), Fundamentals of Future Information
    Technology, 52425 J\"{u}lich, Germany}
    \affil[5]{Diamond Light Source Ltd., Harwell Science and Innovation Campus, Didcot, Oxfordshire, OX11 0DE, United Kingdom}
    \affil[6]{Experimentalphysik IV A, RWTH Aachen University, 52074 Aachen, Germany}
    \affil[$\ddagger$]{\textit{Corresponding author E-Mail:} c.kumpf@fz-juelich.de}
    \affil[$\dagger$]{\textit{Corresponding author E-Mail:} fabian.goehler@physik.tu-chemnitz.de}
    \affil[*]{\textit{Corresponding author E-Mail:} philip.schädlich@physik.tu-chemnitz.de}

\begin{document}

	
\maketitle

 \par

\begin{abstract}

    Quantum Spin Hall Insulators (QSHI) have been extensively studied both theoretically and experimentally because they exhibit robust helical edge states driven by spin-orbit coupling and offer the potential for applications in spintronics through dissipationless spin transport. However, to realise devices, it is indispensable to gain control over the interaction of the active layer with the substrate, and to protect it from environmental influences. Here we show that a single layer of elemental Bi, formed by intercalation of an epitaxial graphene buffer layer on SiC(0001), is a promising candidate for a QSHI. 
    This layer can be reversibly switched between an electronically inactive precursor state and a ``bismuthene state'', the latter exhibiting the predicted band structure of a true two-dimensional bismuthene layer. 
    Switching is accomplished by hydrogenation (dehydrogenation) of the sample, i.e., a partial passivation (activation) of dangling bonds of the SiC substrate, causing a lateral shift of Bi atoms involving a change of the adsorption site. In the bismuthene state, the Bi honeycomb layer is a prospective QSHI, inherently protected by the graphene sheet above and the H-passivated substrate below. Thus, our results represent an important step towards protected QSHI systems beyond graphene.
	
\end{abstract}


	
    Emerging quantum materials, distinguished by different forms of strongly correlated electrons in solids, may pave the way towards next-generation applications such as dissipationless quantum transport or quantum computing \cite{TokuraNat.Phys.2017}.
    In two-dimensional (2D) Quantum Spin Hall (QSH) materials, topologically protected spin-polarised edge states exist at zero magnetic fields \cite{RothScience2009}.
    As proposed by Kane and Mele \cite{KanePhys.Rev.Lett.2005, KanePhys.Rev.Lett.2005a}, graphene represents such a material, however, with an extremely small topological band gap in the sub-meV range. In contrast, a single honeycomb layer of heavy atoms, such as Bi \cite{LiuPhys.Rev.Lett.2011, MurakamiPhys.Rev.Lett.2006}, is considered to be an ideal Quantum Spin Hall Insulator (QSHI) due to the strong spin-orbit coupling that results in a substantial bulk band gap \cite{HsuNewJ.Phys.2015, ZhouProc.Natl.Acad.Sci.2014}.  
    Such honeycomb layers have been realised experimentally, e.g., in the form of indenene, which can be considered as a "hidden" honeycomb despite its triangular structure \cite{BauernfeindNat.Commun.2021}, and bismuthene \cite{ReisScience2017} on SiC, where non-trivial gaps can open due to spin-orbit coupling and the orbital filtering effect \cite{Zhou_2014, HsuNewJ.Phys.2015, ZhouProc.Natl.Acad.Sci.2014}, an effect that moves the $p_{z}$ orbital away from the Fermi level due to hybridisation with the underlying substrate's dangling bonds. However, thin metal films are problematic for use in device applications due to their sensitivity to the environment from which they need to be protected. 
	
    The intercalation of epitaxial graphene on SiC is a well suited way to produce such protected metallic films. Known as confinement heteroepitaxy \cite{BriggsNat.Mater.2020}, this technique has emerged as a promising way to stabilise new 2D quantum materials such as 2D gallium, indium, tin, and lead \cite{BriggsNat.Mater.2020, SchmittNat.Commun.2024, Kim_2016, SchaedlichAdv.Mater.Interfaces2023}.
    In this process, the intercalant moves to the interface, decouples the graphene-like buffer layer (zeroth-layer graphene, ZLG) from the substrate and transforms it into a quasi-freestanding graphene (QFG) layer. The QFG layer subsequently protects the intercalated layer against environmental degradation \cite{SchmittNat.Commun.2024}. 
    Intercalation of Bi under the buffer layer has been the subject of previous studies, but the initial focus was primarily on its effect on the graphene properties \cite{StoehrPhys.Rev.B2016,SohnJ.KoreanPhys.Soc.2021,Wolff_2024}.	
    For the Bi layer, however, a honeycomb lattice was predicted to be the energetically most favourable structure at a coverage of 2/3 of a monolayer \cite{HsuSurf.Sci.2013}. 
    And indeed, such a layer can be obtained experimentally by Bi deposition and subsequent annealing \cite{SohnJ.KoreanPhys.Soc.2021,Wolff_2024}. Here we demonstrate that this Bi layer is a precursor phase for 2D bismuthene, and that it can be reversibly transformed into bismuthene by hydrogenation. 
    Moreover, we explain the mechanism driving the phase transition at the atomistic level and present 
    the electronic and the geometric structures of both phases. In the bismuthene state, the Bi layer exhibits all ingredients for a 2D QSH material.


\section*{Electronic and Geometric Properties of the Bismuthene Precursor Phase}

\begin{figure}[t!]
        \centering
        \includegraphics[scale=1]{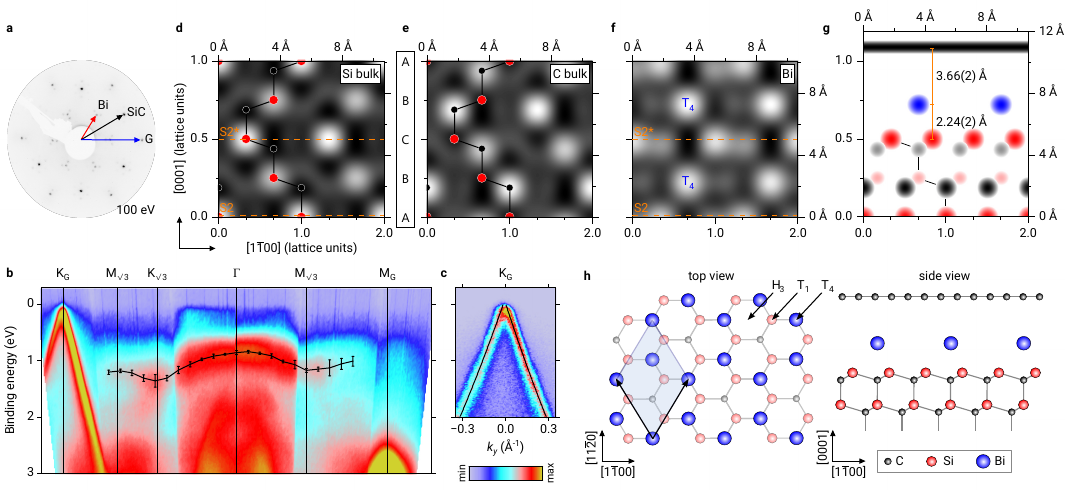}
        \caption{\textbf{Electronic and geometric structure of the bismuthene precursor phase:}
        \textbf{a} LEED pattern with the reciprocal lattice vectors of graphene (G), SiC, and Bi highlighted ($E =$ \qty{100}{eV}). 
        \textbf{b} Energy-momentum cuts along the $\Gamma-\text{K}_{\text{G}}$ and $\Gamma-\text{M}_{\text{G}}$ directions of graphene. The weakly dispersing low-energy state originating from the bismuthene precursor is marked by a black line derived from fitting the corresponding energy distribution curves.
        \textbf{c} Energy-momentum cut at K$_{\text{G}}$ in the direction perpendicular to the map shown in \textbf{b}. The linear dispersion of the \textpi-band is indicated in black, as obtained by fitting the maxima of the momentum distribution curves using a nearest-neighbour tight-binding approximation for graphene.
        \textbf{d}-\textbf{f} NIXSW imaging results for bulk-Si (\textbf{d}), bulk-C (\textbf{e}), and Bi (\textbf{f}). The Fourier-reconstructed electron densities in a plane spanned by the [1$\overline{\text{1}}$00] and [0001] crystallographic directions are shown, white maxima correspond to the positions of the atoms. A ball-and-stick model of the bulk structure is superimposed on \textbf{d} and \textbf{e} and fits to the experimental finding very well. In \textbf{f}, the clearest maxima are located at $\text{T}_\text{4}$ sites, indicating the adsorption sites of Bi. Orange dashed lines in \textbf{d} and \textbf{f} indicate the two possible SiC surface terminations S2 and S2* (A and C planes).
        \textbf{g} Structure of the graphene-protected bismuthene precursor sample, as derived from NIXSW imaging. The maxima from maps \textbf{d}, \textbf{e} and \textbf{f} are shown in red, gray and blue, indicating the positions of bulk-Si, bulk-C and Bi, respectively. Note that in order to mimic the ball-and-stick model from \textbf{h}, maxima from the panels \textbf{d}-\textbf{f} and from another adjacent parallel plane, offset by half a lattice vector, are shown for the bulk species. Graphene, since it is incommensurate with the bulk structure, is shown as a horizontal black bar at its corresponding adsorption height.
        \textbf{h} Top view and side view of the proposed atomic arrangement for the bismuthene precursor structure on SiC(0001) in a ($\sqrt{3} \times \sqrt{3}$)$R30^{\circ}$ superstructure. Bi is located on the $\text{T}_\text{4}$ hollow sites, while the $\text{H}_\text{3}$ hollow and $\text{T}_\text{1}$ on-top sites are unoccupied. The graphene layer has been omitted in the top view for clarity.
        }
        \label{fig:1}
    \end{figure}

    We start with a discussion of the properties of the bismuthene precursor phase, which is prepared by Bi intercalation of a graphene buffer layer and several subsequent annealing steps (see Methods section for details). It has been designated ``Bi \textbeta\ phase'' in recent literature \cite{SohnJ.KoreanPhys.Soc.2021, Wolff_2024}, since its potential for transformation into 2D bismuthene has not been fully recognised before. Low energy electron diffraction (LEED, see \autoref{fig:1}\textbf{a}) reveals a very well ordered layer system with ($\sqrt{3} \times \sqrt{3}$)$R30^{\circ}$-periodicity. 
    The electronic structure as determined by angle-resolved photoelectron spectroscopy (ARPES) is shown in \autoref{fig:1}\textbf{b}. The \textpi-bands of graphene are clearly observed with the Dirac cone at the K$_{\text{G}}$ point, indicating that the graphene is well decoupled from the substrate and the Bi layer below. \autoref{fig:1}\textbf{c} shows a cut through the K$_{\text{G}}$ point in the plane perpendicular to that shown in panel \textbf{b}. The Bi layer, however, shows no signatures that would be characteristic of the 2D material bismuthene, but rather a weakly dispersing band about \qtyrange{1.0}{1.5}{eV} below the Fermi energy $E_{\textrm{F}}$, as indicated by the black line in \autoref{fig:1}\textbf{b} marking the maxima of this band. This will change when the transformation from precursor to bismuthene is performed, as discussed below.

    To date, the structure of this phase has not been solved unambiguously, although the question of the Bi adsorption site in particular is a crucial one. It determines the bonding configuration between the Bi atoms and the substrate, and thus also within the Bi layer. We have addressed this question by applying an extension of the Normal Incidence X-ray Standing Waves (NIXSW) technique \cite{Zegenhagen2013,Woodruff_2005}, known as NIXSW imaging \cite{bedzyk2004}. It is based on a Fourier component analysis, which provides the element-specific density distribution of all atomic species in the unit cell, and since it is not susceptible to the phase problem of diffraction methods, it is able to resolve the full three-dimensional structure of the bulk and the surface. Details of the technique are given in the Methods section. A full discussion of the analysis of this data set is published elsewhere \cite{ImagingPaperPlaceHolder}.

    \autoref{fig:1}\textbf{d}-\textbf{f} shows the results for the bismuthene precursor phase. 2D cuts through the three-dimensional atomic density distribution for the bulk species Si and C are shown in \textbf{d} and \textbf{e}, respectively, and clearly resemble the well-known bulk crystal structure of 4H-SiC, as illustrated by the superimposed ball-and-stick model. However, the most relevant result, i.e., the density distribution for Bi shown in \autoref{fig:1}\textbf{f}, is more difficult to interpret, as it involves two different surface terminations: It is known that 4H-SiC with its atomic plane stacking sequence A-B-C-B-A undergoes step bunching when annealed for surface preparation, because the A- and C-terminated terraces are energetically favoured over the B termination \cite{PakdehiAdv.Funct.Mater.2020}. Atomic force microscopy results shown in Sec.\ 1 of the Supplementary Material indicate that this is also the case for our samples; we find a balanced distribution of the so-called S2 (A-plane) and S2* (C-plane) surface terminations. In \autoref{fig:1}\textbf{d} these two terminations are indicated by the horizontal, orange dashed lines, marking the positions of the uppermost Si atoms (surface planes). We also indicate these planes in panel \textbf{f}, because the Bi atomic positions, as obtained by NIXSW imaging, have to be evaluated relative to one of these planes. We find two maxima in the Bi density distribution (marked by $\text{T}_\text{4}$) which are located at reasonable vertical distances to one of the surface planes, and therefore represent Bi adsorption sites. Their vertical distance to the surface plane is \qty{2.24}{\angstrom}, smaller than the expected length of a Si-Bi covalent bond (\qty{2.67}{\angstrom} \cite{Pyykkö_2009}), but consistent with a covalent bond between Si and a Bi atom on a hollow site. The lateral position of the maxima in \autoref{fig:1}\textbf{f} indicates that Bi is located above the C atom of the terminating SiC bilayer (compare with \autoref{fig:1}\textbf{e}), that is the $\text{T}_\text{4}$ adsorption site. Note that the other (weaker) maxima in the map do not appear at reasonable distances from either of the two surface planes. The strongest of these, if interpreted as an adsorption site, would correspond to the $\text{T}_\text{4}$ site above a B-terminated surface, which may even be present on the surface to some small extent. However, it is more likely that the smaller maxima are artefacts due to the finite sum of Fourier components (finite number of different Bragg reflections) used in the experiment (see Methods).
    
    We conclude that in the bismuthene precursor phase the Bi atoms adsorb exclusively on $\text{T}_\text{4}$ sites. \autoref{fig:1}\textbf{g} shows the complete structure for the example of a S2* (C-plane) terminated substrate: 
    All maxima (above a certain threshold) of the maps (\textbf{d}-\textbf{f}) are shown in one plot, colour-coded red, grey, and blue for Si, C, and Bi, respectively. The position of the graphene layer above Bi is also shown here, as a black bar \qty{3.66}{\angstrom} above the Bi layer. No lateral structure can be resolved for graphene since it is incommensurate with the bulk lattice. However, the vertical distance between Bi and graphene is close to the sum of the van der Waals radii of the two species involved (\qty{3.77}{\angstrom} \cite{Mantina_2009}), indicating that the Bi-graphene interaction is almost exclusively van der Waals-like.
    Note that our finding of Bi in $\text{T}_\text{4}$ hollow sites, although clearly evidenced by NIXSW imaging, is in contrast to the model discussed by Sohn et al. \cite{SohnJ.KoreanPhys.Soc.2021} who suggested the other hollow site of the SiC surface, $\text{H}_\text{3}$ (see \autoref{fig:1}\textbf{h}, top view). However, our finding is consistent with theoretical predictions for the most favourable adsorption sites on the Si-rich ($\sqrt{3} \times \sqrt{3}$)$R30^{\circ}$ reconstruction of SiC(0001) \cite{NorthrupPhys.Rev.B1995, SabischPhys.Rev.B1997}. These authors identified $\text{T}_\text{4}$ as the most favourable site, owing to strain relaxation in the uppermost SiC bulk layer. The two C atoms below the Bi adatoms relax downwards, increasing their bond angles closer to the ideal tetrahedral angle of \qty{109.5}{\degree}, while the third C atom in the ($\sqrt{3} \times \sqrt{3}$)$R30^{\circ}$ supercell, which has no adatom above it, shifts upwards. This type of strain relief is not possible in the $\text{H}_\text{3}$ configuration, where the uppermost C bulk atoms are equivalent \cite{NorthrupPhys.Rev.B1995, SabischPhys.Rev.B1997}. 

    The adsorption of Bi on a hollow site has decisive consequences for its bonding configuration. As can be seen from the ball-and-stick model shown in \autoref{fig:1}\textbf{h} (top view), 
    on the $\text{T}_\text{4}$ site each Bi atom has three equidistantly located Si neighbours, and the Bi layer therefore saturates all Si dangling bonds. Note that we have drawn a Bi honeycomb here, since the results obtained for bismuthene (see below) are suggestive of this structure. 
    However, since NIXSW imaging is restricted to the bulk unit cell, we cannot unambiguously discriminate equivalent $\text{T}_\text{4}$ sites within the ($\sqrt{3} \times \sqrt{3}$) superstructure cell. But adsorption sites other than $\text{T}_\text{4}$ can be excluded, and therefore each Bi atom forms bonds with three Si atoms, inhibiting a planar hybridization and thus the formation of a Bi layer of true 2D character. As we will demonstrate in the following, this is changed by a hydrogen-induced transition from the precursor phase to a true 2D bismuthene phase.

\section*{Geometric Structure of the 2D Bismuthene phase}

\begin{figure}[t!]
        \centering
        \includegraphics[scale=1]{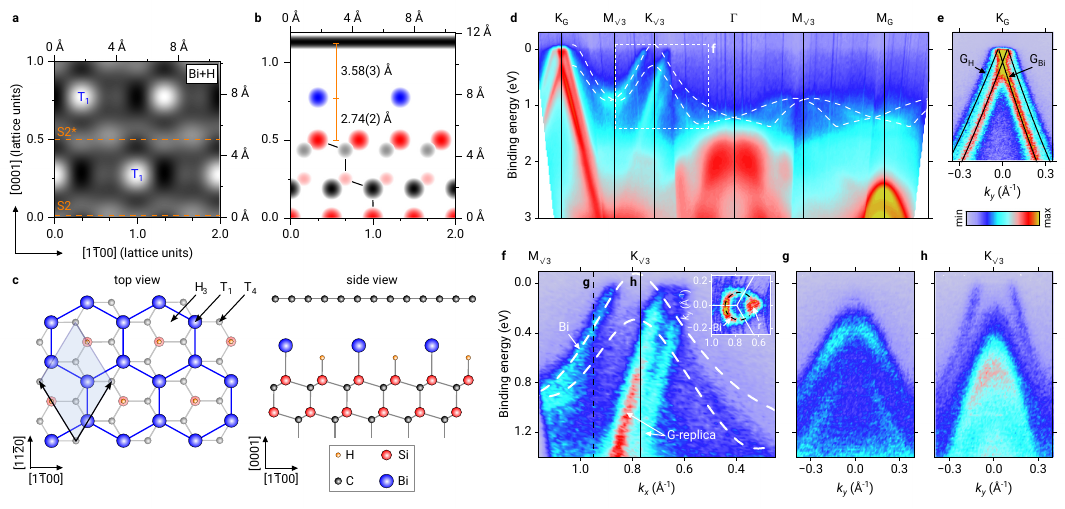}
        \caption{\textbf{Geometric and electronic structure of the 2D bismuthene layer on SiC(0001) (after hydrogenation):} 
        \textbf{a} NIXSW imaging result for Bi, shown for the plane spanned by the [1$\overline{\text{1}}$00] and [0001] crystallographic directions. White maxima correspond to the positions of the atoms. 
        \textbf{b} Structure of the graphene-protected bismuthene sample, as derived from NIXSW imaging. Red, gray, and blue circles represent the maxima of the Fourier-reconstructed electron densities for bulk-Si, bulk-C and Bi, respectively. Graphene, since it is incommensurate with the bulk structure, is shown as a horizontal black bar at its corresponding adsorption height. 
        \textbf{c} Top view and side view of a ball-and-stick model showing the determined arrangement of the bismuthene honeycomb on the SiC(0001) surface in a ($\sqrt{3} \times \sqrt{3}$)$R30^{\circ}$ superlattice. The graphene layer has been omitted in the top view for clarity. 
        \textbf{d} Energy-momentum band maps along the $\Gamma-\text{K}_{\text{G}}$ and $\Gamma-\text{M}_{\text{G}}$ directions of graphene. The white dashed lines represent the valence band structure of bismuthene as calculated by DFT. 
        \textbf{e} Energy-momentum map around K$_{\text{G}}$ of graphene in the direction perpendicular to \textbf{d}. Two energetically shifted Dirac cones of graphene are observed, due to the coexistence of graphene areas on bismuthene ($\text{G}_{\text{Bi}}$) and hydrogen intercalated graphene areas ($\text{G}_{\text{H}}$). Black lines were obtained by fitting the maxima of momentum distribution curves with the nearest-neighbor tight-binding band structure of graphene. 
        \textbf{f} Close-up in the vicinity of K$_{\sqrt{3}}$ as marked in \textbf{d} by a white dashed rectangle. The inset shows the Fermi contour of the bismuthene in a $k_x$-$k_y$ map. Brillouin zone boundaries are drawn as white lines. The left part of the contour, indicated by a black dashed line, corresponds to the bismuthene contribution, in the right part it is superimposed by the graphene replica cones (marked ``r''). 
        \textbf{g}, \textbf{h} $E$-$k_y$ energy-momentum maps in the direction perpendicular to the map shown in \textbf{f}, at the position indicated by a black vertical dashed line in \textbf{f} and at the K$_{\sqrt{3}}$ point of bismuthene, respectively.
         }
        \label{fig:2}
    \end{figure}

    The transformation of the precursor phase into a 2D bismuthene phase is performed by hydrogenation in a two-step annealing process in a H$_2$ atmosphere (see Methods section for details). After hydrogenation we have again performed NIXSW imaging, ARPES and X-ray photoelectron spectroscopy (XPS) measurements.
    In Figures \ref{fig:2}\textbf{a} and \textbf{b} we show the NIXSW imaging results for the sample after hydrogenation. While the results for the bulk species are unchanged (not shown), a comparison of Fig.\ \ref{fig:2}\textbf{a} (after hydrogenation) with \ref{fig:1}\textbf{f} (before hydrogenation) demonstrates that the position of the Bi atoms has changed, i.e., the atoms have rearranged throughout the Bi layer (see also Supplementary Information Sec.\ 2). \autoref{fig:2}\textbf{b} shows the complete structure in the combined plot (similar to \autoref{fig:1}\textbf{g}), also illustrated by the corresponding ball-and-stick model in \autoref{fig:2}\textbf{c} (to be compared with \autoref{fig:1}\textbf{h}).
    These results reveal two important aspects: 
    (i) Bi has moved away from the $\text{T}_\text{4}$ hollow site, which it occupied in the precursor phase. In bismuthene, it occupies the $\text{T}_\text{1}$ site on top of the uppermost Si atoms. 
    (ii) The Bi layer did not decouple completely from the substrate by the hydrogenation process, but the Bi atoms are now located almost precisely at the distance of a covalent Si-Bi bond, at \qty{2.74}{\angstrom} above the topmost Si atoms. (The sum of the covalent radii of the two species is \qty{2.67}{\angstrom} \cite{Pyykkö_2009}.) 
    Thus, our structure determination indicates a vertically oriented, covalent bond between the Bi atoms on $\text{T}_\text{1}$ sites and the topmost Si atoms underneath, a result consistent with the structural model proposed by Reis et al.\ for bismuthene on a H-saturated SiC(0001) surface \cite{ReisScience2017} (a system not covered by graphene). 
    In addition, the Bi-graphene separation of \qty{3.58}{\angstrom} reveals van der Waals interactions only between Bi and the overlying protective graphene layer.

    The Bi adsorption site determines the bonding configuration also within the bismuthene layer. In fact, the change of the adsorption site from $\text{T}_\text{4}$ in the precursor phase to $\text{T}_\text{1}$ in bismuthene is the key for understanding the phase transition and the 2D properties of bismuthene. In the precursor phase, each Bi atom saturates three substrate dangling bonds in the unit cell, see above. But the existence of three bonds to underlying Si atoms does not allow for the planar hybridisation of Bi that is needed for the formation of the characteristic band structure of a 2D bismuthene honeycomb. After hydrogenation, when Bi adsorbes on $\text{T}_\text{1}$, only one covalent Bi-Si bond is formed, involving the $p_z$ orbital and one of the five valence electrons of Bi. The remaining $s$ and $p$ orbitals with four valence electrons allow for a planar hybridization within the Bi layer. Thus, this orbital filtering effect removes exactly one orbital (and valence electron) from the in-plane binding configuration. It is required to establish the formation of the distinct Dirac-like bands being characteristic for 2D bismuthene, but only possible when Bi is adsorbed on a $\text{T}_\text{1}$ on-top position above Si.  
          
\section*{Electronic Properties of 2D Bismuthene Underneath Epitaxial Graphene}

    The proof that the hydrogenated Bi layer is indeed a true 2D bismuthene layer comes from the determination of the electronic structure by means of ARPES. The results of these measurements are shown in \autoref{fig:2}\textbf{d}-\textbf{h}. From the $E$-$k_x$ energy-momentum band map shown in \autoref{fig:2}\textbf{d} it is evident that the weakly dispersing band discussed above for the non-hydrogenated precursor phase has vanished and instead strongly dispersing Dirac-like bands emerge close to $E_{\textrm{F}}$ between the M$_{\sqrt{3}}$ and the K$_{\sqrt{3}}$ points. 
    In the close-up shown in \autoref{fig:2}\textbf{f} these bands are even more clearly visible. The white dashed lines represent density functional theory (DFT) calculations for the QSHI state of bismuthene (see Supplementary Material, Sec.\ 3 for details). 
    The steep bands next to them, labeled ``G-replica'' and crossing the Fermi level just to the right of the K$_{\sqrt{3}}$ point, are replicas of the graphene Dirac cones, backfolded due to the SiC lattice \cite{PolleyPhys.Rev.B2019} and partly covering the bismuthene bands to the left of the K$_{\sqrt{3}}$ point.
    The upper one of the bismuthene calculated bands (upper white-dashed line)  matches very well with the left of the Dirac-like bands visible in the ARPES data (marked ``Bi'' in \autoref{fig:2}\textbf{f}), while the lower band is better to be seen in an $E$-$k_y$-map perpendicular to the maps shown in panels \textbf{d} and \textbf{f}. \autoref{fig:2}\textbf{g} shows such a perpendicular map through the position marked by the black dashed line labelled ``g'' in \textbf{f}. The splitting of these upper and lower Dirac-like bands, best to be seen in \autoref{fig:2}\textbf{g}, is a Rashba-type spin splitting, as demonstrated by the DFT calculations detailed in Sec.\ 3 of the Supplementary Information, and is thus ascribed to the inversion symmetry breaking (ISB) at the interface. 
    We also mention that the peak shapes of XPS measurements confirm that the bismuthene phase is metallic, in contrast to the precursor phase which is insulating (see Supplementary Material, Sec.\ 4 for details).
    

    Another fingerprint of a QSHI state is a band gap in the Dirac cone at the Fermi edge, i.e., a deviation from the linear dispersion expected for honeycomb lattices with negligible spin-orbit coupling. Such a gap is not directly visible in the data shown in \autoref{fig:2}, which is due to a significant p-doping of our bismuthene layer that shifts the bands upwards (in contrast to Ref.\ \cite{ReisScience2017}). It is known from the literature that the p-doping is due to the pyroelectric nature of the substrate, where a residual bulk dipole moment persists, the magnitude of which depends on the SiC polytype \cite{QteishPhys.Rev.B1992}. The p-doping of the bismuthene layer is well visible in the inset of \autoref{fig:2}\textbf{f}, which shows a $k_x$-$k_y$ section around K$_{\sqrt{3}}$ directly at the Fermi surface. The circular contour on the left side (black dashed line) is a horizontal cut through a Dirac cone which is still of finite size at the Fermi level. On the right, the ring is superimposed by a signal from the graphene replica cone (labelled ``r''). At the surface, this spontaneous polarisation acts as a pseudo-acceptor layer, generating a negative sheet of charge density that induces holes in adjacent materials \cite{RisteinPhys.Rev.Lett.2012, Mammadov2DMater.2014}. For our case, the p-doping can be quantified as $p_\text{Bi}=\text{\qty{1.84(0.12)e13}{\cm^{-2}}}$.
    
    \begin{figure}[t!]
        \centering
        \includegraphics[scale=1]{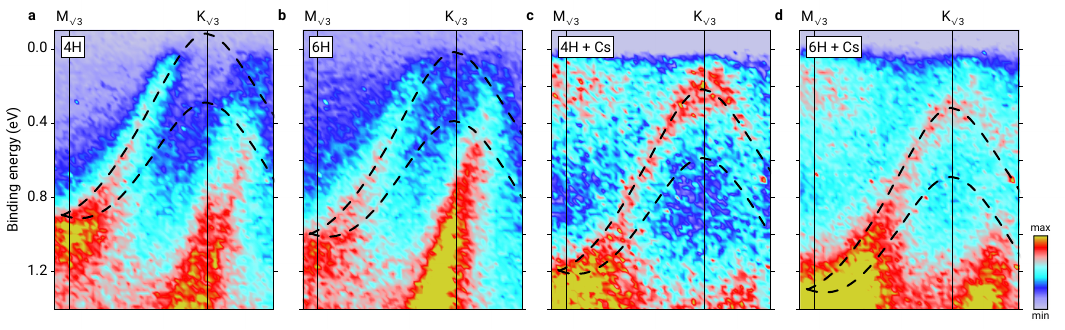}
        \caption{\textbf{Energy-momentum band maps of bismuthene at different doping levels.} Black dashed lines indicate the DFT calculated dispersion of the Bi Dirac cone. A rigid band shift was applied to the calculated bands to match the experimental results. Different doping levels were obtained by the use of different substrate polytypes 4H-SiC and 6H-SiC, which imply  different spontaneous polarisation strengths, and by additional adsorption of the alkali metal Cs on top of graphene, inducing a charge transfer into the bismuthene layer. \textbf{a} 4H-SiC substrate, no Cs. \textbf{b} 6H-SiC substrate, no Cs. \textbf{c} 4H-SiC substrate, with Cs. \textbf{d} 6H-SiC substrate, with Cs. In \textbf{d} the valence band maximum is positioned significantly below the Fermi level, indicating the presence of a band gap.}
        \label{fig:5}
    \end{figure}

    While this p-doping makes it impossible to observe the gap in the Dirac cone directly, it can be made visible in ARPES measurements performed at different doping levels, as presented in \autoref{fig:5}. 
    These are energy-momentum maps along the K$_{\sqrt{3}}-$M$_{\sqrt{3}}$ direction, similar to \autoref{fig:2}\textbf{f}, but for different substrates and with Cs adsorbed additionally on top of the graphene. Both modifications cause downshifts of the band structure of different magnitude, due to (i) the 6H polytype exhibiting a weaker spontaneous polarisation \cite{Mammadov2DMater.2014}, and hence reducing the charge carrier concentration ($p_\text{Bi}=\text{\qty{4.1(1.5)e12}{\cm^{-2}}}$) 
    compared to the 4H polytype, and (ii) due to the alkali atoms providing some additional n-doping of the bismuthene without intercalating the Bi or graphene layer \cite{Goehler2DMater.2024}. The downshift of these two measures is $\approx$ \qty{0.1}{eV} due to the 6H substrate, and $\approx$ \qty{0.3}{eV} due to Cs adsorption. 
    The series of energy-momentum maps in \autoref{fig:5}\textbf{a} to \textbf{d}, measured for different polytypes and with or without Cs adsorption, shows a stepwise downshift of the bismuthene band structure. While in \textbf{a} and \textbf{b} (4H-SiC and 6H-SiC, respectively, no Cs), the valence band maximum is still above or near the Fermi level, in \textbf{c} and \textbf{d} (both substrates with Cs) it moves further down, clearly demonstrating the existence of a band gap.

    Finally, it should be mentioned that the ARPES data shown in \autoref{fig:2} also contains the band structure of graphene. Two sets of \textpi-bands can be observed at the K$_{\text{G}}$ point, shifted in energy with respect to each other, as apparent in the $E$-$k_y$-map shown in \autoref{fig:2}\textbf{e}. One comes from the graphene layer on bismuthene, the other from areas that are not Bi-intercalated. Such areas arise during the preparation of the precursor phase. Before hydrogenation, in these areas only the graphene buffer layer (ZLG) is present, which in the hydrogenation process turns into H-intercalated QFG. Since QFG and bismuthene areas have different doping levels, two Dirac cones arise. Before hydrogenation, we observe only one single Dirac cone (from the graphene on top of the precursor phase, see \autoref{fig:1}\textbf{c}, with the Dirac point located only slightly above the Fermi energy, indicating that this graphene layer is almost undoped.


    We conclude from the experimentally obtained band structure of the hydrogenated Bi layer and its very good agreement with the DFT calculations -- in particular from the existence of the Dirac-like band, its spin splitting, and the energy gap at the Fermi edge -- that this Bi layer is indeed a single layer of 2D bismuthene. For the first time, a bismuthene layer was grown underneath a graphene layer and is thus protected from the environment. 

\section*{Reversibility of the Precursor to Bismuthene Phase Transition}

    It is known that a purely H-intercalated QFG layer is only stable up to \qty{700}{\degreeCelsius} in ultra-high vacuum (UHV) \cite{RiedlPhys.Rev.Lett.2009}. The formation of our bismuthene phase was carried out well below this temperature, by annealing at \qty{550}{\degreeCelsius} in H$_2$ atmosphere (see Methods). Thus, the bismuthene is expected to be not stable above \qty{700}{\degreeCelsius}. In Sec.\ 5 of the Supplementary Material, in particular in Fig.\ S.6(e), we show that indeed the transition from the precursor phase to the bismuthene phase is reversed by annealing at \qty{700}{\degreeCelsius} in UHV. The electronic bands associated with bismuthene have disappeared and the weakly dispersing band, characteristic of the precursor phase, has reestablished. 
    It is remarkable that both the precursor-to-bismuthene transformation by hydrogenation, as well as the bismuthene-to-precursor transformation by annealing (and dehydrogenation) are reversible. We have performed the hydrogenation -- dehydrogenation cycle several times on one sample, and always found largely the same band structures in ARPES for both precursor and bismuthene phases. Thus, the 2D properties of bismuthene can be switched on and off by hydrogenation and dehydrogenation. 
    
    A common feature of both phases is the ($\sqrt{3} \times \sqrt{3}$)$R30^{\circ}$ LEED pattern. This reconstruction, which indicates a Bi coverage significantly below $1$~ML, appears to be an important ingredient of the precursor phase as well. This is also reflected by the fact that other intercalated Bi phases, e.g., the \textalpha\ phase consisting of a close-packed unreconstructed layer with ($1 \times 1$) perodicity, cannot be converted into bismuthene, as discussed in Sec.\ 6 of the Supplementary Information. This can be understood by the mechanism behind the phase transition:  The H-saturation of $1/3$ of the Si dangling bonds, which in turn triggers the change of the Bi adsorption site from $\text{T}_\text{4}$ to $\text{T}_\text{1}$, is only possible when the Bi coverage is (locally) at or below $2/3$ of a monolayer. At higher coverages, e.g., in the case of the \textalpha\ phase, all Si dangling bonds are saturated one-by-one by Bi atoms \cite{Wolff_2024}, which inhibits the H-saturation and thus the formation of the honeycomb structure of bismuthene.
    


\section*{Conclusion}

    In this study, we report the formation of a 2D bismuthene honeycomb layer, a prospective quantum spin Hall insulator, sandwiched between a graphene layer above and the SiC substrate below. The evidence for having a true 2D bismuthene layer is the excellent agreement of our ARPES data with DFT results for the QSHI state of bismuthene (own calculations and Refs.\ \cite{ReisScience2017, HsuSurf.Sci.2013, HsuNewJ.Phys.2015}). In particular, the archetypal Dirac-like bands near $E_\textrm{F}$, a band gap around the K$_{\sqrt{3}}$ point, and a pronounced Rashba-type energy splitting of the bismuthene valence bands are clearly revealed in ARPES.

    The bismuthene is formed in a reversible transition from a precursor, the ($\sqrt{3} \times \sqrt{3}$)$R30^{\circ}$ reconstructed intercalated Bi \textbeta\ phase. The phase transition is driven by hydrogenation and involves a lateral movement of the Bi atoms from the $\text{T}_\text{4}$ hollow adsorption site to the $\text{T}_\text{1}$ site on top of the uppermost Si bulk atoms. This change in adsorption site is key to understanding the phase transition as it determines the bonding configuration in the Bi layer: In the precursor phase each Bi atom has three equidistant Si neighbours, in the bismuthene phase only one. Only in the latter case, the distinct Dirac-like bands can form, since exactly one orbital (the $p_z$ orbital) is excluded from the in-plane hybridisation. This orbital filtering is indispensable for the formation of the characteristic band structure of the 2D bismuthene honeycomb. 
    
    Stability is obviously a crucial aspect for any epitaxial 2D (multi-)layer system, in particular for future applications. Due to the encapsulation by the graphene layer above and the SiC substrate below, the bismuthene layer proved to be stable in air. In Sec.\ 5 of the Supplementary Material we demonstrate that the band structure of the graphene/bismuthene/SiC system has not changed after 24 hours of exposure to air. The bismuthene is indeed efficiently protected from the environment. Thus, we have successfully fabricated a switchable, air-stable bismuthene QSHI with a large band gap and strong Rashba spin splitting. These results represent an important step forward on the road towards future QSH devices.
    

\section*{Methods}	
	
\subsection*{Sample preparation}

    Epitaxial zeroth-layer graphene (ZLG) substrates were prepared using the polymer-assisted sublimation growth on SiC described in detail elsewhere \cite{Kruskopf2DMater.2016}, a method that is well-known to provide graphene samples with superior quality compared to those grown in ultra high vacuum (UHV). SiC wafers were purchased from Pam-Xiamen. 
    Bi intercalation was carried out following a deposition and annealing approach reported previously \cite{SohnJ.KoreanPhys.Soc.2021,Wolff_2024}: 
    After degassing the samples in UHV at \qty{450}{\degreeCelsius}, Bi was deposited on the SiC surface from a custom-built Knudsen cell (operated at a temperature of \qty{550}{\degreeCelsius}) for \qty{120}{min} to evaporate a thin layer of elemental Bi. Deposition took place in a dedicated chamber with a base pressure better than \qty{5e{-9}}{mbar}.
    After in-vacuo transfer to the UHV analysis system the samples were annealed at \qty{450}{\degreeCelsius} for \qty{30}{min}, which leads to Bi intercalation of the ZLG and the formation of the \textalpha\ phase, a densely packed Bi layer underneath graphene. 
    
    Subsequently the bismuthene precursor phase (Bi \textbeta\ phase) was produced by annealing the \textalpha\ phase sample at \qty{950}{\degreeCelsius}, leading to a partial depletion of Bi from the intercalation layer and the formation of a ($\sqrt{3} \times \sqrt{3}$)$R30^{\circ}$ superlattice, as verified by LEED (see \autoref{fig:1}\textbf{a}) \cite{Wolff_2024, SohnJ.KoreanPhys.Soc.2021}. The sample temperature during annealing was controlled by a pyrometer, assuming a sample emissivity of 0.9. After a fast transport through air (no longer than \qty{5}{min}), the hydrogenation of the samples was carried out by exposure to ultra-pure H$_2$ (\qty{880}{mbar}, \qty{0.9}{slm}) in a dedicated tube furnace \cite{SeyllerJ.Phys.:Condens.Matter2004}. Samples were first heated to \qty{300}{\degreeCelsius} for \qty{10}{min}, followed by the main hydrogenation process at \qty{550}{\degreeCelsius} for \qty{90}{min}. This leeds to the formation of the bismuthene phase in all regions of the sample surface that were Bi intercalated. In some cases, H-intercalated quasi-freestanding graphene was also formed in small domains due to an overall lack of Bi, since then ZLG domains are formed in the precursor phase.
  
\subsection*{Photoelectron spectroscopy \& Low-energy electron diffraction}	
	 
    After growth, the samples were transfered in vacuum into a UHV chamber dedicated for X-ray photoelectron spectroscopy (XPS), angle-resolved photoelectron spectroscopy (ARPES), and low-energy electron diffraction (LEED). The chamber was equipped with a monochromatized SPECS Focus 500 \ce{Al-K\textalpha} X-ray source, a monochromatized SPECS UVS 300 ultraviolet light source providing linear polarized \ce{He}-I and \ce{He}-II radiation, and a SPECS Phoibos 150 hemispherical electron analyzer with 2D-CCD detector. For LEED, a SPECS ErLEED 150 was used. The base pressure of the UHV system was better than \qty{3e-10}{mbar}.
    
\subsection*{Normal incidence X-ray standing waves (NIXSW) and NIXSW imaging}     

    For the normal incidence X-ray standing waves experiments, the samples were brought to beamline I09 of the Diamond Light Source in Didcot, UK. A dedicated UHV transport chamber (``vacuum suitcase'') was used for the transport, which allowed transfer in UHV both in Chemnitz, where the samples were grown and pre-characterized, and at the Diamond beamline. The chamber was operated at a pressure better than \qty{5e-10}{mbar}. The data was measured in the UHV system of the NIXSW endstation of beamline I09 using a Scienta EW4000 HAXPES analyser. In such an NIXSW experiment, in general, XPS data is recorded from the sample surface while an X-ray standing wave field, generated by the interference of an incident and a Bragg reflected X-ray wave, is established in the bulk crystal and at its surface 
    \cite{Zegenhagen2013,Woodruff_2005,Zegenhagen1993,bocquet2019torricelli}. Given the well-defined position of the standing wave with respect to the crystal lattice, the photoelectron yield of all atomic species depends on the positions of the atoms in the unit cell. The method is therefore capable of localizing any ensemble of absorber atoms with respect to the Bragg planes of the reflection used to generate the standing wave field. It is most commonly used to measure the vertical position of adsorbates on surfaces or interlayer distances above crystalline substrates (see e.g., Refs.\ \cite{Wolff_2024, SchaedlichAdv.Mater.Interfaces2023, LinPhys.Rev.B2022,vanStraatenJElSpec.2018,StadlerNatPhys.2009,StadtmuellerJElSpec.2015,KleinJPCC.2019}, and references in \cite{Zegenhagen2013,Woodruff_2005}). In this standard case, a Bragg reflection with a scattering vector perpendicular to the surface must be chosen. The technique provides separate structural information for all those species in the sample, the photoemission signals of which can be separated.

    The NIXSW-based Fourier imaging technique (NIXSW imaging) that we applied in this work goes a decisive step further: It exploits the fact that the result of one NIXSW measurement with any chosen Bragg reflection $\boldsymbol{H}=(hkl)$ represents the $\boldsymbol{H}$th Fourier component of the atomic density of the atomic species under consideration.  Thus, collecting data using a sufficient number of different non-equivalent reflections allows the reconstruction of the element-specific atomic density relative to the bulk unit cell. In this way, the geometric structure of the sample (both bulk and surface) can be derived using the well-established formalism of inverse Fourier transforms. Therefore, this technique is not susceptible to the general phase retrieval problem inherent in diffraction techniques, as the NIXSW method provides both amplitude and phase information.    
    Although this technique was first proposed about two decades ago \cite{bedzyk2004} , it has not been widely used until now due to its significant technical requirements. We apply it here, for the first time to our knowledge, to a 2D material system. We performed NIXSW measurements on seven inequivalent Bragg reflections using the core levels C\,1s, Si\,2s and Bi\,4f$_\text{7/2}$. In the C\,1s spectra, the signal from the graphene and bulk carbon can be separated easily due a large core level shift. The analysis of the data sets leading to the results reported here is explained in detail elsewhere \cite{ImagingPaperPlaceHolder}.

\subsection*{Computational Details}

    Density functional theory (DFT) as implemented in the ABINIT package \cite{gonze2009abinit} was used to calculate the structural and electronic properties of the systems. For this purpose, fully relativistic norm-conserving pseudopotentials with the generalized gradient approximation (GGA) in the Perdew-Burke-Ernzerhof (PBE) \cite{perdew1996generalized} formulation were used to evaluate the exchange-correlation potential. Optimization of the initial structures were performed using a $k$-point grid of $8\times 8\times 1$ until residual forces smaller than $5\times10^{-6}$ eV/Å were achieved. For an accurate ground state description, a $12\times 12\times 1$ grid was adopted. In all cases, a 1088 eV cutoff energy was used for the plane wave basis set and spin-orbit coupling was considered. Spin-textures were plotted using the PyProcar \cite{pyprocar} package.  
    In our calculations, the graphene layer was not included in order to reduce the computational cost. Hence, we focus on the bismuthene phase on SiC. The dangling bonds at the lower carbon surface of SiC were passivated with H atoms and a vacuum layer of 15 {\AA} was used.


\section*{Author contributions}	

    N.T., C.K., T.S., and P.S.\ conceived the project. 
    The sample preparation was carried out by N.T.
    The PES and LEED measurements at TU Chemnitz were performed and analyzed by N.T.\ and P.S.
    The NIXSW experiments were performed by N.T, S.W., S.S., T.-L.L., C.K., and analyzed by N.T.\ and S.W., with significant input from S.S.\ and C.K.
    A.D.P.U.\ and S.G.\ performed and interpreted the DFT calculations.  
    N.T., F.S.T., C.K., T.S., and P.S. developed the atomistic model explaining the reversible phase transition. 
    All authors discussed the results. N.T.\ made the figures, and N.T., C.K., F.G., and P.S.\ wrote the paper, with significant input from all authors.


\section*{Acknowledgments}	
	
    The authors would like to thank Christoph Lohse for substrate preparations. We also thank the Diamond Light Source for access to beamline I09 through proposal SI36085-2, and the I09 beamline staff (Pardeep Kumar Thakur and Dave McCue) for technical support. This work was supported by the German Research Foundation (Deutsche Forschungsgemeinschaft, DFG) within the Research Unit FOR5242 (project 449119662) and the Collaborative Research Centre SFB-1083 (project A12).


\begin{singlespace}

\end{singlespace}

\end{document}


\maketitle



\section{Step structure of the 4H-SiC substrate}

\begin{figure}[h!]
        \centering
        \includegraphics[scale=1]{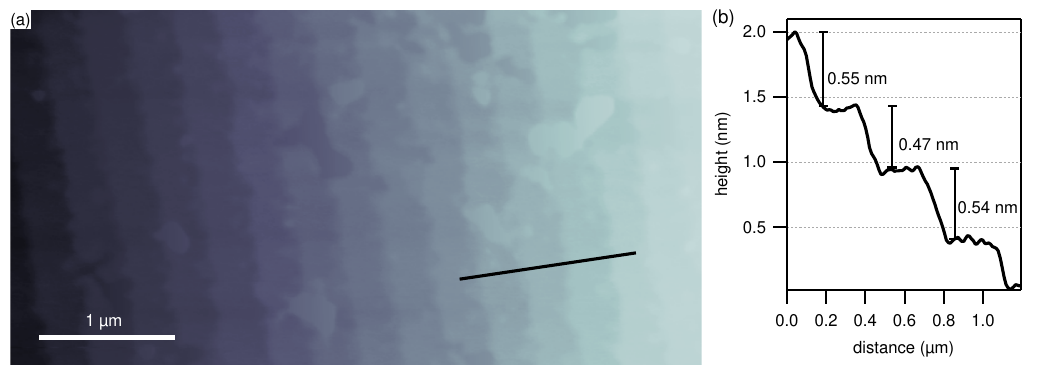}
        \caption{(a) AFM image showing the topography of a buffer layer (zeroth layer graphene) on a 4H-SiC(0001) surface before Bi intercalation. (b) Height profile measured along the black line in (a).}
        \label{fig:AFM}
    \end{figure}

    In \autoref{fig:AFM}(a) we show an atomic force microscopy (AFM) image of a graphene buffer layer sample on 4H-SiC prior to Bi intercalation. From left to right, a sequence of step edges can be seen, separating flat terraces that are about \qty{0.3}{\textmu m} wide. The height profile in (b), recorded along the black line shown in (a), indicates a typical step height of $\sim$ \qty{0.5}{nm}, a value corresponding to the height of two SiC bilayers, that is half of the 4H-SiC unit cell. This allows for either S1/S1* or S2/S2* terminated terraces, but since the first is energetically unfavourable \cite{PakdehiAdv.Funct.Mater.2020}, we conclude that the vast majority of the surface is S2 and S2* terminated.
    
\newpage
\section{Low-energy electron diffraction: I(V) analysis of the precursor and the bismuthene phase}

     \begin{figure}[h!]
        \centering
        \includegraphics[scale=1.8]{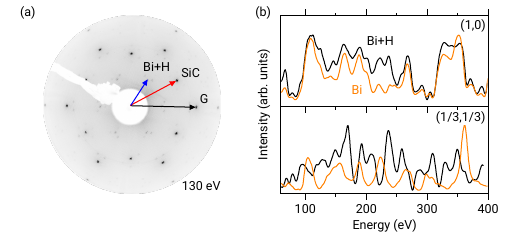}
        \caption{(a) LEED pattern (E = 130 eV) with selected reciprocal lattice vectors of graphene (G), SiC, and bismuthene (Bi+H) highlighted. (b) LEED-IV spectra for selected beams before (Bi, orange) and after (Bi+H, black) the hydrogenation process.}
        \label{fig:placeholder}
    \end{figure}
    
   Evidence for a significant structural rearrangement of the Bi layer upon hydrogenation can be obtained from low-energy electron diffraction. \autoref{fig:placeholder}(a) shows the LEED pattern obtained for the bismuthene phase. At this electron energy, the pattern is very similar to that of the precursor phase (Figure 1\textbf{a} of the main text). In particular, the ($\sqrt{3} \times \sqrt{3}$) superstructure associated with the Bi layer (blue arrows labeled Bi+H) is robust throughout the hydrogenation process. However, the LEED-IV spectra shown in \autoref{fig:placeholder}(b) for one bulk reflection (the (10), red arrow in (a)) and one reflection from the Bi layer (the ($\frac{1}{3} \frac{1}{3}$), blue arrow) reveal clear differences in the intensities of the reflections for the two phases, in particular for the second reflection. We performed a Pendry analysis, known as a way to quantify the relevant differences in LEED-IV spectra \cite{Pendry_1980}. For the SiC bulk spots we obtain $R_{p}=0.24$, for the first order diffraction spots of the ($\sqrt{3} \times \sqrt{3}$) superstructure  $R_{p}=1.03$. The large value obtained for the latter is a clear indication of fundamental structural changes in the interfacial structure, as associated with a change of the Bi adsorption site from $\text{T}_\text{4}$ to $\text{T}_\text{1}$.

\newpage

\section{Theoretical model and spin texture}
     
     \begin{figure}[h!]
        \centering
        \includegraphics[scale=1]{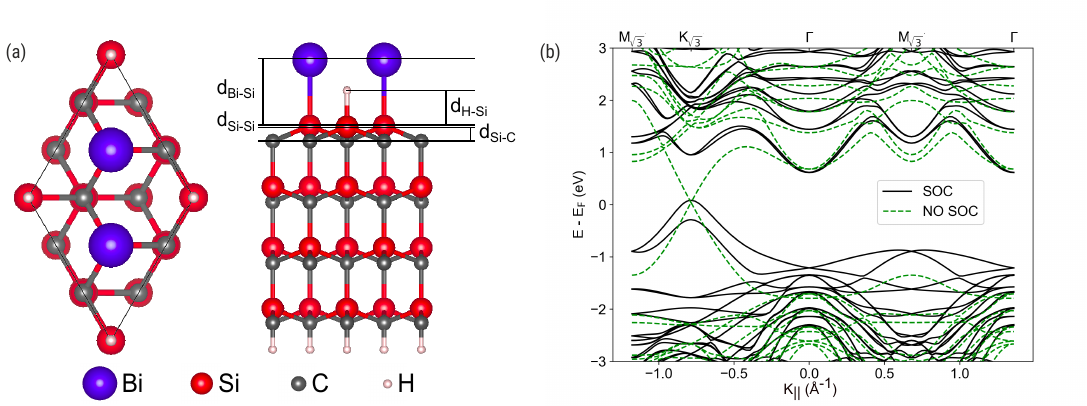}
        \caption{(a) Atomic structure of the bismuthene phase on 4H-SiC after hydrogenation (left: top view, right: side view onto the (10$\overline{\text{1}}$0) plane of SiC). (b) Corresponding calculated band structure. Black solid lines and green dashed lines represent calculations with and without consideration of the spin-orbit coupling, respectively.}
        \label{fig:DFT}
    \end{figure}     

In \autoref{fig:DFT} (a), a ball and stick model of bismuthene on SiC (after hydrogenation) is presented.  In our calculations, the details of which can be found in the methods section of the main text, a lattice constant of \SI{5.33}{\angstrom} was used. After performing structural relaxation, the obtained distance between Bi and Si atoms was $d_{\text{Bi-Si}}=\SI{2.75}{\angstrom}$, which is in excellent agreement with the experimental value found by NIXSW imaging ($\SI{2.74}{\angstrom}$, see the main text). For the hydrogen atom located on top of the top-most Si atom, a Si-H distance of $\SI{1.5}{\angstrom}$ was obtained. Furthermore, the top-most Si plane is found slightly buckled ($d_{\text{Si-Si}}=\SI{0.034}{\angstrom}$) and a vertical separation between the Si and C atoms $d_{\text{Si-C}}=\SI{0.64}{\angstrom}$ ($\SI{0.606}{\angstrom}$ for the Si atom saturated with H) was observed. 

\begin{figure}[h!]
        \centering
        \includegraphics[scale=1.2]{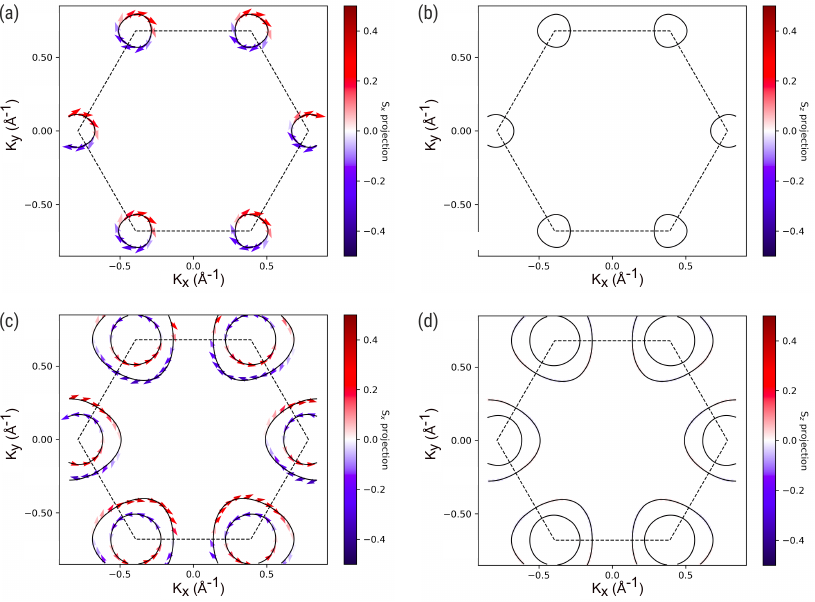}
        \caption{   Calculated spin texture for bismuthene (a,b) at $E_{\textrm{F}}$ and (c,d) at $E_{\textrm{B}}=0.5\, \textrm{eV}$. (a) and (c) represent an in-plane projections ($S_x$), (b and d) out-of-plane projections ($S_z$). 
        }
        \label{fig:Spin}
    \end{figure}

In \autoref{fig:DFT} (b) the calculated electronic structure is displayed. Black solid lines and green dashed lines represent calculations with and without taking into account spin-orbit coupling (SOC), respectively. Note that a Dirac-like dispersion appears at the K point in the case that SOC was not included. Once SOC is included in the calculations, the two corresponding bands split with a gap of \SI{0.9}{\eV} and a global indirect band gap of \SI{0.54}{\eV} is obtained. 
Moreover, if SOC was included in the calculation, a Rashba-type spin splitting of \SI{0.37}{\eV} is found in the top valence band at the K point, as shown in \autoref{fig:Spin}.

~

To confirm the Rashba splitting of the top valence band induced by the broken inversion symmetry of the bismuthene layer, 2D constant energy contours of the spin texture were calculated in the $k_{x}-k_{y}$ plane centered at the $\Gamma$ point. \autoref{fig:Spin} shows the projected $S_{x}$ and $S_{z}$ spin components on the constant energy contours at (a,b) $E_{F}$ and (c,d) $E_{B}=\SI{0.5}{\eV}$. 
In (a) we observe a radial spin polarization for the upper branch of the top valence band around the six $K$ points. The second branch, which is visible at $E_{B}=\SI{0.5}{\eV}$ (panel (c)), shows the opposite spin polarisation. 
In both cases, however, the out-of-plane $S_{z}$ component is negligible, as seen in (b) and (d), indicating a complete in-plane spin polarization as it is characteristic of Rashba-type spin splitting.

\newpage

~

\newpage
\section{X-ray photoelectron spectroscopy analysis of core levels}
\begin{figure}[h!]
        \centering
        \includegraphics[scale=1.35]{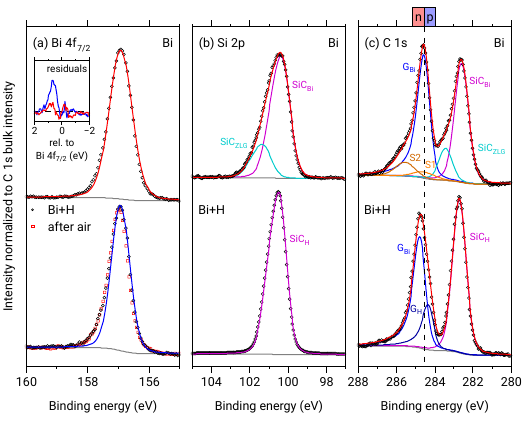}
        \caption{(a-c) X-ray photoelectron spectra of the Bi\,4f$_{7/2}$, Si\,2p and C\,1s core levels, respectively, for the precursor (top) and the bismuthene phase (bottom). In (a), the two spectra (black data points) were fitted by symmetric line profiles (solid lines). The residuals shown in the inset indicate a change in the line profile upon hydrogenation. The spectrum plotted as red data points in the bottom of panel (a) was recorded after exposure of the sample to air. 
        The Si\,2p (b) and C\,1s spectra (c) were fitted using multiple components, see text. The vertical, dashed line in (c) indicates the C\,1s peak position for charge neutral graphene.}
        \label{fig:4}
    \end{figure}

    In Figures \ref{fig:4}(a)-(c) we show representative X-ray photoelectron spectra of the precursor phase and the bismuthene phase. Upon hydrogenation the shape of the Bi\,4f peak changes from symmetric to asymmetric, as can be seen in panel (a): While the Bi\,4f peak before hydrogenation (upper spectrum labelled Bi) can be described by a symmetric line profile, a similar fit after hydrogenation (lower spectrum, black data points labelled Bi+H) fails, as also indicated by the residuals of both fits shown in the inset. This change in spectral shape is in agreement with the transition from an insulating state in the initial precursor phase to a p-doped metallic state of the bismuthene seen in the ARPES measurements (see main text). The ratio of the (integrated) intensities of Bi\,4f and the Si\,2p is barely affected by hydrogenation, suggesting a stable overall coverage of Bi. In conjunction with the LEED diffraction patterns, this suggests a similar in-plane arrangement of the Bi atoms. 
    Note that the Bi\,4f signal does not change after exposing the sample to ambient conditions for \qty{24}{h} (\autoref{fig:4}(a), bottom, red data points), see also Sec.\ 5.

    The Si\,2p and C\,1s spectra (see Figures \ref{fig:4} (b) and (c)) have more complicated profiles due to more complex bonding environments of the involved atomic species. Before hydrogenation, the bulk contributions of the spectra consist of two components (SiC\sub{Bi} and SiC\sub{ZLG}), indicating slight de-intercalation owing to the preparation of the precursor phase from the \textalpha\ phase by annealing (see also Methods section). The components are separated because of a different surface band bending underneath intercalated and non-intercalated regions. The degree of intercalation is about \qtyrange{70}{80}{\percent}, with the non-intercalated part of the sample surface being covered by a ZLG. Hence, in the C\,1s spectrum, not only the signal (G\sub{Bi}) stemming from the graphene layer on top of the precursor phase but also the components S1 and S2 attributed to the ZLG are identified.
    
    After H-intercalation we observe significant changes in the Si\,2p and C\,1s spectra. The graphene C\,1s peak now consists of two components, while the ZLG components have vanished. This is due to the ZLG being converted into H-intercalated QFG, which gives rise for a G\sub{H} component close to the G\sub{Bi} peak. Thus, beside the graphene-protected bismuthene (originating from the precursor domains) we now find regions of H-intercalated QFG on the surface. The C\,1s peak positions indicate slight n- and p-type doping of the respective graphene layers, in agreement with the ARPES results, see main text.    
    The Si\,2p spectra can now be fitted by only one single component, which indicates that both of these domains exhibit a comparable surface band bending.

\newpage
\section{Stability in air and de-hydrogenation}



    \begin{figure}[b!]
        \centering
        \includegraphics[scale=1]{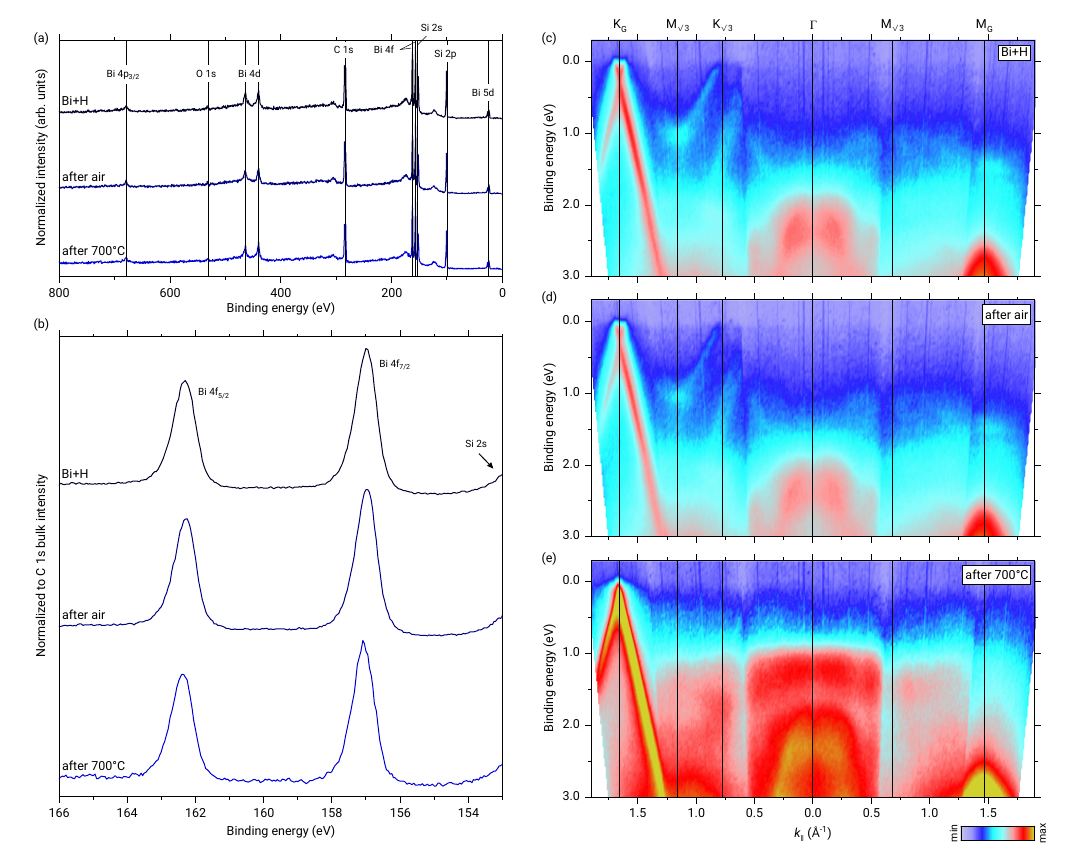}
        \caption{PES data of a bismuthene sample directly after preparation (Bi+H), after exposure to air for \qty{24}{h} (after air), and after subsequent annealing at \qty{700}{\degreeCelsius} (after \qty{700}{\degreeCelsius}). (a) XP survey spectra, indicating that no significant oxidation takes place. 
        (b) Bi\,4f core level spectra (normalized to the respective C\,1s bulk peak). (c-d) ARPES data taken along the $\text{K}_{\text{G}}-\Gamma-\text{M}_{\text{G}}$ direction of graphene, indicating that the band structure of graphene is preserved during exposure to air.
        }
        \label{fig:air}
    \end{figure}

    As discussed in the main text, a key feature of the system presented in this work is the intrinsic protection of the quantum spin Hall insulator bismuthene against environmental degradation provided by the graphene sheet above (and the substrate below) the Bi layer. In order to check the efficiency of this protection, we exposed a bismuthene sample to air for \qty{24}{h}. The upper overview spectrum shown in \autoref{fig:air}(a) was recorded from a bismuthene sample (Bi+H) immediately after preparation, revealing that no significant amount of oxygen is present on the sample. After exposure to air, this has not changed, as demonstrated by the spectrum shown in the center of the panel labelled ``after air'', confirming that no oxidation took place at the surface and in the heterolayer system.     
    This conclusion is confirmed by the XP spectra of the Bi\,4f core level shown in \autoref{fig:air}(b). The shape of the Bi\,4f core level spectra recorded before and after exposure to air (Bi+H and after air) shown in the upper part and the center of the panel are basically unchanged, in particular, no additional shifted peaks can be observed that could be attributed to oxidized bismuth. For reference, the binding energy of the Bi\,4f$_{7/2}$ orbital in $\text{Bi}_2 \text{O}_3$ is \qty{158.7}{eV} \cite{zhong1994bi}.

    However, the proof for the bismuthene layer being stable under ambient conditions is provided by ARPES. The photoemission maps presented in \autoref{fig:air}(c) and (d) depict the band structures of the sample immediately after preparation and after exposure to air for \qty{24}{h}, respectively. Also here, no significant differences between the two spectra can be observed, confirming the stability of the graphene / bismuthene heterostructure in air. 

    Furthermore, we have performed experiments to investigate the thermal stability of the system. The limiting factor for thermal stability is expected to be the H-saturation of the Si dangling bonds, i.e., the Si-H bonds located in the center of the bismuthene honeycomb. Previous studies on hydrogen intercalation of epitaxial graphene reported de-intercalation, i.e., the breaking of the Si-H bond, at annealing temperatures of approximately \qty{700}{\degreeCelsius} \cite{RiedlPhys.Rev.Lett.2009}. A similar stability is therefore expected for the present system. The lower curves in \autoref{fig:air} (a and b) display the survey XP spectrum and the Bi\,4f core level spectrum, respectively, after annealing at \qty{700}{\degreeCelsius}. No significant changes are observed when comparing the two spectra to those before annealing. In particular, the intensities of the Bi\,4f spectra (which are normalized to the C1s bulk intensity), reveal that the overall Bi coverage remains unchanged.

    The ARPES data obtained for the sample after annealing is shown in  \autoref{fig:air} (e). It clearly demonstrates that due to the annaling the band structure has reverted to that of the precursor phase (before hydrogenation). In particular, the characteristic bismuthene Dirac bands have vanished, and the insulating state previously associated with the precursor phase has been reestablished. Furthermore, the doping of the graphene layer has changed from n-type to charge neutrality, consistent with the observations for the precursor phase.

    We have performed the hydrogenation - dehydrogenation cycle several times on the same sample with the same results, and thus conclude that both the hydrogenation of the precursor phase and the dehydrogenation of the bismuthene are reversible processes.

\newpage

\section{Hydrogenation of the \textalpha\ phase}
    
    \begin{figure}[h!]
        \centering
        \includegraphics[scale=1]{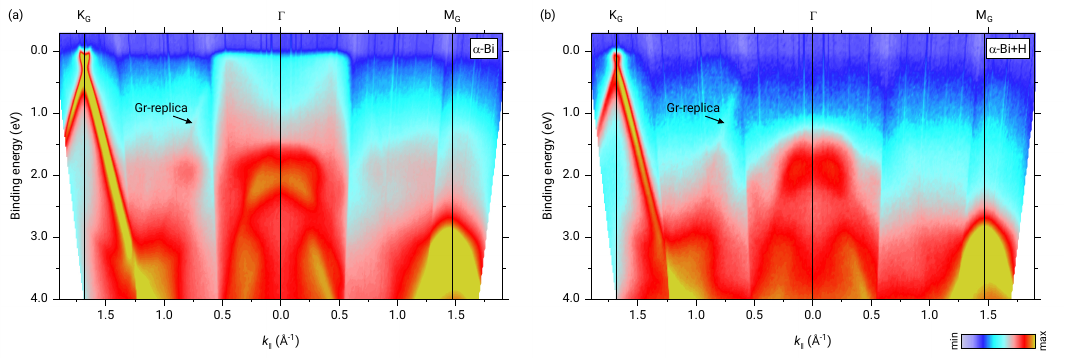}
        \caption{ARPES data of a Bi \textalpha\ phase (\textalpha-Bi) along $\text{K}_{\text{G}}-\Gamma-\text{M}_{\text{G}}$ direction of graphene before (a) and after (b) hydrogen treatment. Both photoemission maps show no indications for the formation of bismuthene.}
        \label{fig:alpha}
    \end{figure}

    \autoref{fig:alpha} shows photoemission maps of the Bi \textalpha\ phase before (a) and after (b) hydrogen treatment. No significant changes of the electronic structure are observed. In particular, no bismuthene bands can be observed in \autoref{fig:alpha} (b), but only the valence band maximum of the SiC substrate at $\Gamma$ and the graphene \textpi-bands with their characteristic linear dispersion at K$_{\mathrm{G}}$ together with a \textpi-band replica. 
    
    We conclude that beside the saturation with hydrogen, also the structural $\left(\sqrt{3} \times \sqrt{3}\right)$ template of the Bi \textbeta\ structure is essential for the formation of bismuthene.


\begin{singlespace}

\end{singlespace}